\documentclass[preprint,amsmath,amssymb,superscriptaddress]{revtex4-1}

\usepackage{graphicx}
\usepackage{dcolumn}
\usepackage{bm}
\usepackage[utf8]{inputenc}
\usepackage[T1]{fontenc}
\usepackage{mathptmx}
\setcitestyle{square}
\usepackage{color}

\usepackage{booktabs} 

\usepackage{nicefrac} 

\usepackage{amsfonts}

\usepackage{lipsum}
\usepackage{wasysym}

\begin{document}
\pagecolor{white}

\title{Ultra-short, MeV-scale laser-plasma positron source for positron annihilation lifetime spectroscopy}

\author{Thomas L.~Audet}
\affiliation{Centre for Plasma Physics,
  School of Mathematics and Physics,
  Queen's University Belfast
 , BT7 1NN, Belfast United Kingdom}
 
 \author{Aaron ~Alejo}
\affiliation{Centre for Plasma Physics,
  School of Mathematics and Physics,
  Queen's University Belfast
 , BT7 1NN, Belfast United Kingdom}
 
 \author{Luke Calvin} 
\affiliation{Centre for Plasma Physics,
  School of Mathematics and Physics,
  Queen's University Belfast
 , BT7 1NN, Belfast United Kingdom}
 
 \author{Mark Hugh Cunningham}
\affiliation{Centre for Plasma Physics,
  School of Mathematics and Physics,
  Queen's University Belfast
 , BT7 1NN, Belfast United Kingdom}
 
 \author{Glenn Ross ~Frazer}
\affiliation{Centre for Plasma Physics,
  School of Mathematics and Physics,
  Queen's University Belfast
 , BT7 1NN, Belfast United Kingdom}
 
 \author{Nasr A. M. ~Hafz}
 \affiliation{ELI-ALPS. ELI-HU Non-profit Ltd., H-6728 Szeged, Hungary}
 \affiliation{National Laboratory on High Power Laser and Physics, SIOM, CAS, Shanghai 201800, China}
 \affiliation{Dept. of Plasma and Nuclear Fusion, Nuclear Research Center, Atomic Energy Authority, Abu-Zabal 13759, Egypt}
 
  \author{Christos ~Kamperidis}
 \affiliation{ELI-ALPS. ELI-HU Non-profit Ltd., H-6728 Szeged, Hungary}
 
  \author{Song ~Li}
 \affiliation{ELI-ALPS. ELI-HU Non-profit Ltd., H-6728 Szeged, Hungary}
  
\author{Gagik Nersisyan}
\affiliation{Centre for Plasma Physics,
  School of Mathematics and Physics,
  Queen's University Belfast
 , BT7 1NN, Belfast United Kingdom}
  
\author{Daniel ~Papp}
 \affiliation{ELI-ALPS. ELI-HU Non-profit Ltd., H-6728 Szeged, Hungary}
 
\author{Michael ~Phipps}
\affiliation{Centre for Plasma Physics,
  School of Mathematics and Physics,
  Queen's University Belfast
 , BT7 1NN, Belfast United Kingdom}

 \author{Jonathan Richard ~Warwick}
\affiliation{Centre for Plasma Physics,
  School of Mathematics and Physics,
  Queen's University Belfast
 , BT7 1NN, Belfast United Kingdom}

\author{Gianluca ~Sarri}
 \email{g.sarri@qub.ac.uk} 
\affiliation{Centre for Plasma Physics,
  School of Mathematics and Physics,
  Queen's University Belfast
 , BT7 1NN, Belfast United Kingdom}

\begin{abstract}
Sub-micron defects represent a well-known fundamental problem in manufacturing since they can significantly affect performance and lifetime of virtually any high-value component. Positron annihilation lifetime spectroscopy is arguably the only established method capable of detecting defects down to the sub-nanometer scale but, to date, it only works for surface studies, and with limited resolution. Here, we experimentally and numerically show that laser-driven systems can overcome these well-known limitations, by generating ultra-short positron beams with a kinetic energy tuneable from 500 keV up to 2 MeV and a number of positrons per shot in a 50 keV energy slice \color{black} of the order of $10^3$. Numerical simulations of the expected performance of a typical mJ-scale kHz laser demonstrate the possibility of generating MeV-scale narrow-band and ultra-short positron beams with a flux exceeding $10^5$ positrons/s, of interest for fast volumetric scanning of materials at high resolution.

\end{abstract}


\maketitle

\section{Introduction}

Positron Annihilation Lifetime Spectroscopy (PALS) \cite{krause1999positron} is arguably one of the most successful techniques for the non-invasive inspection of materials and identification of small-scale defects. PALS presents several unique advantages when compared to other inspection techniques: it works virtually with any type of material (crystalline and amorphous, organic and inorganic, biotic and abiotic), it can identify even sub-nanometer defects with concentrations as low as less than a part per million, and it can provide information on the type of defect and its characteristic size \cite{krause1999positron,charltonbook}. PALS has found applications in testing systems as diverse as turbines, polymers, semiconducting devices, biomimetic systems, zeolites, and solar cells. 

Even small-scale defects can have a dramatic effect on the performance and lifetime of high-performance and high-value components, especially when made in, and required to perform under, hostile environments. Heat and pressure treatments, new welding methods, radiation exposure, impact damage, are all examples of scenarios that can leave sub-micron defects in materials during advanced manufacturing or extreme performance use. 

In a nutshell, PALS relies on the temporally resolved detection of gamma-rays resulting from the annihilation of positrons as they interact with the material \cite{Keeble:2010aa}. In a perfect crystal lattice, an implanted positron would rapidly thermalize and subsequently annihilate from a delocalised state. However, a positron is likely to be trapped in the potential induced by a vacancy, such as a missing atomic core \cite{PhysRevB.41.9980}. A trapped positron will thus have a more localised state and, therefore, a longer lifetime. The temporal evolution of the gamma-ray emission from the material will thus contain several exponential decays, each with a typical timescale characteristic of the bulk material and of any defects in it. Normally, positron lifetimes in materials are of the order of 200 ps, with longer lifetimes if defects are present (see, for instance, Ref. \cite{Keeble:2010aa}). 

Typical machines designed for PALS routinely operate at a positron energy in the keV range and bunch durations of the order of hundreds of picosecond (see, for instance, Refs. \cite{sponsor,nepomuc,pleps,NANOPOS,PULSTAR}). Despite the high performance of these machines and their wide use for industrial applications and fundamental science, they mainly suffer of two well-known limitations.
First, the available positron energy restricts material scanning only to micron-scale depths. Second, the positron bunch duration is relatively long and thus affects the resolution of the technique. For higher resolution, it is preferable to have positron bunch durations that are significantly smaller than the timescales of interest, i.e., at least in the range of a few to tens of ps. In that case, the resolution of the system will be only limited by the detector response.

These limitations can be overcome if laser-driven positrons are used. Commercially available high power lasers with short pulse durations (fs to ps), can routinely generate high-charge relativistic electron beams of similar duration \cite{RevModPhys.81.1229}. Positrons can then be obtained as the result of the electromagnetic cascade initiated by these relativistic electrons as they propagate through a high Z converter target.  For sufficiently thick converters, two mechanisms are mainly involved: the emission of a high energy photon through bremsstrahlung and the subsequent decay of the photon in electron-positron pairs, with both processes mediated by the nuclear field. Further cascading is also possible, but it is unlikely for thicknesses shorter than a radiation length \cite{Sarri:2013aa}. Several works (see, for instance, Refs. \cite{Sarri:2013aa,Sarri_2013,Sarri:2015aa,PhysRevLett.102.105001,Sarri_2016,Chen:2009aa,PhysRevLett.105.015003,PhysRevLett.114.215001,Sunn_Pedersen_2012,Liang:2015aa,Lieaav7940}) have already reported on the generation of positrons from laser-driven electron beams, mainly following two approaches based on whether the electrons are generated during direct laser irradiation of the converter target or if they are first generated in a gaseous medium following, for instance, the laser-wakefield acceleration (LWFA) mechanism \cite{RevModPhys.81.1229}.



In this article, we demonstrate that positron beams with characteristics appealing to PALS can be produced in a compact configuration, using laser-driven electron beams. A simple and compact beam-line consisting of two Hallbach quadrupole magnets, an energy selector, and two dipole magnets is already capable of generating relatively high fluxes of positrons per second, with a duration of the order of tens of picoseconds and a kinetic energy seamlessly tuneable from approximately 0.5 to 2 MeV. The performance of the system is extracted from Monte-Carlo simulations and validated in a proof-of-principle experiment using the  TARANIS laser hosted by the Centre for Plasma Physics at Queen's University Belfast \cite{Dzelzainis:2010aa}. The extension of these results to the use of a kHz mJ-level laser system, such as SYLOS2 \cite{Toth:2020aa} at ELI-ALPS, indicates that more than $10^5$ MeV-scale positrons per second in a 50 keV energy slice can be generated, with a duration of the order of 90 ps, i.e., shorter than the typical timescales of positron annihilation in materials \cite{Keeble:2010aa}. This work thus provides a proof-of-principle validation for a laser-based compact positron source of interest for industrial applications as proposed, for instance, in the conceptual design report of the EuPRAXIA plasma-based accelerator \cite{EuPRAXIA}.\color{black}

In Sec.~\ref{sec:DirectPositrons} we will describe the main experimental and numerical results concerning positron beam generation and transport using a direct laser-solid interaction scheme. In Sec~\ref{sec:Extension} we will show numerical results of extending this work to high repetition rate low-energy laser systems. A final discussion and concluding remarks will then be provided in Sec.~\ref{sec:Conclusion}.

\section{Positron generation by direct laser solid irradiation}
\label{sec:DirectPositrons}

\subsection{Experimental setup}
\label{sec:ExperimentalSetup}

\begin{figure}[b!]
\centering
\vspace{5mm}
\includegraphics[width=1\textwidth]{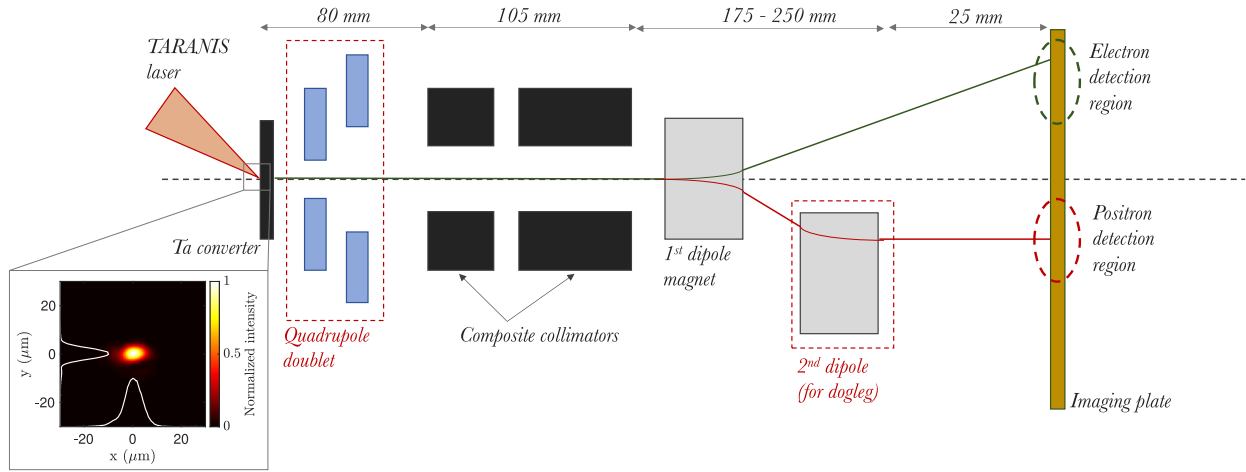}
\caption{Sketch of the experimental setup. Three different configurations were considered: \textit{full system} (including the quadrupole doublet and the second dipole), \textit{collimation only} (with quadrupole doublet but no second dipole), \textit{generation only} (no quadrupole doublet and no second dipole). Additionally, some shots have been taken with a thin gold target (50 $\mu$m) to measure and characterize the initial electron spectrum starting the positron generation inside the converter.}
\label{fig:configurationSketches}
\vspace{5mm}
\end{figure}

We first discuss the experimental results of tests performed using the TARANIS laser facility at the Queen's University Belfast \cite{Dzelzainis:2010aa}. A sketch of the setup is shown in Fig. \ref{fig:configurationSketches}.

TARANIS is a chirped pulse amplification (CPA) laser system based on a Ti:Sapphire front-end and a Nd:Glass amplification section. In our experiment, the system delivered laser pulses with an energy of $E_l = (8.9 \pm 0.5)$ J in a $\tau_l = (0.8 \pm 0.1)$ ps full width at half maximum (FWHM) pulse duration. The typical intensity contrast of the laser is $\sim 10^{-7}$ at 1.5 ns before the main pulse. The laser was focused using an F/3 off-axis parabolic mirror (OAP) to a FWHM spot size of $w_x = (7.6 \pm 0.8)$ $\mu$m and $w_y = (4.9 \pm 0.4)$ $\mu$m in the horizontal and vertical directions respectively, leading to a peak intensity on target of $I_L = (2.5 \pm 0.9) \times 10^{19}$ $\mathrm{W/cm^2}$. The inset in Fig ~\ref{fig:configurationSketches} shows the measured intensity distribution of the laser focal spot onto the target. The angle of incidence of the laser beam on the target was $30^{\circ}$. 

\begin{figure}[b!]
\centering
\includegraphics[width=0.4\textwidth]{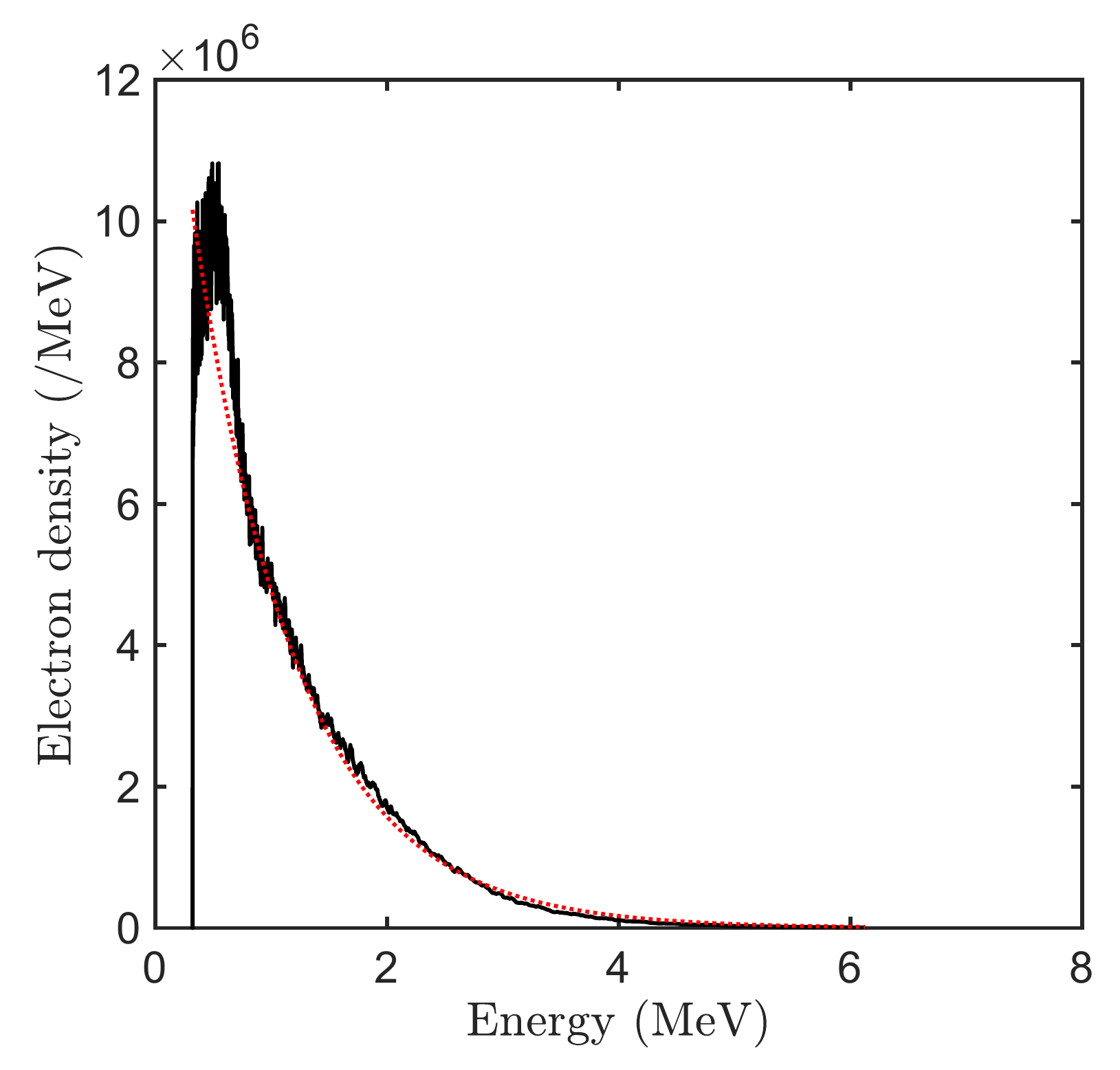}
\caption{Measured electron spectrum (black solid line) and Maxwellian fit (dashed red line) after the interaction of the TARANIS laser (details in the text) with a 50 $\mu$m gold target. 
The Maxwellian fit corresponds to an electron temperature of 0.9 MeV.}
\label{el_spectrum}
\vspace{5mm}
\end{figure}

When focused onto a thin target, the pedestal of the laser pulse generates an overdense plasma with a characteristic keV-scale electron temperature. The interaction of the high-intensity peak of the laser with this cold plasma generates a super-thermal population of electrons, with a characteristic temperature of $T_{hot}\simeq 1$ MeV, which propagates through the target. To experimentally infer the characteristics of this hot electron population, we have first performed a series of preliminary shots on a thin gold target with a thickness of $50\mu$m.
Fig.~\ref{el_spectrum} shows a typical electron spectrum obtained in this part of the experiment. The electron spectrum is well reproduced by a Maxwell-Boltzmann distribution with a temperature of $k_B T_e \simeq 0.9$ MeV, in agreement with the intensity scalings for the $\vec{J}\times \vec{B}$ heating mechanism \cite{Kruer:1985aa}. 
Here, the total number of detected electrons is $N^{e^-}_{detected} \simeq 9.5 \times 10^6$, in a $1.3\times10^{-4}$ steradian cone \color{black}.
 The full cone of emission of the electrons is of the order of 0.7 steradian (consistently with previous work reported in \cite{green}) \color{black}, implying a total number of electrons escaping the rear of the gold target of $N^{e^-}_{emitted} \simeq 5 \times 10^{10}$. This electron population will be considered hereafter as a good approximation for the electron population starting the cascade within a thicker converter target. 

For the rest of the article, we will focus on one specific converter, i.e., a $l_{Ta} = 2$ mm - thick tantalum foil, corresponding to approximately half of a radiation length for that material. It is intended that a different converter material and thickness will generate positron populations with slightly different spectral and spatial qualities. The choice of these parameters must be dictated by the specifics of the laser system and the requirements of the application sought. The choice of 2mm of tantalum is only shown here as an example.  \color{black}

To thoroughly test the performance of the system, different configurations were used after the target, as sketched in Fig.~\ref{fig:configurationSketches}. In all the configurations, two collimators were placed on axis, downstream of the target. The collimators are an assembly of plastic, aluminium and lead. The first collimator consists of a $T_{CH} = 2.25$ mm thick layer of plastic (polyethylene) followed by a $T_{Al} = 5$ mm thick layer of aluminium (Al), followed by a $T_{Pb} = 25$ mm thick layer of lead (Pb). The second collimator consists of a $T_{CH} = 2.25$ mm thick layer of plastic followed by a $T_{Al} = 5$ mm thick layer of Al, followed by a $T_{Pb} = 50$ mm thick layer of Pb. Each collimator has a centered circular aperture with a diameter of $\diameter_1 = 11$ mm and $\diameter_2 = 19$ mm, respectively. 

In the \textit{generation only} configuration, the collimators are followed by a single dipole magnet with an average field of $B = 50$ mT and length of 30 mm. Mapping of the magnetic field distribution inside the dipoles shows a super-Gaussian spatial distribution of the field intensity (index = 4) with a peak field of $B_{max} = 53$ mT and a width $\sigma = 15$ mm. \color{black}. In the \textit{collimation only} configuration, a doublet of quadrupole magnets in the Hallbach configuration (similar to those described in Ref. \cite{PhysRevSTAB.10.082401}) was added in between the target and the collimators to increase the collection and collimation of the positrons. Both quadrupoles are 10 mm long and they are separated by 10 mm. Their inner diameters are 44 mm and 88 mm and their measured \color{black} magnetic field gradients are 17.8 T/m and 8.9 T/m, respectively.
Finally, in the \textit{full system} configuration, a second identical magnetic dipole was added between the first dipole and the detector. This second dipole was placed off-axis, on the positron side to form what is commonly known as a dogleg.

Additional lead shielding (not shown on Fig~\ref{fig:configurationSketches}) was used on each side of the collimators to reduce noise at the detector location. In all configurations, the whole system is in vacuum and the  electrons, positrons and photons were detected using an imaging plate (IP). The IP used was a BAS-SR2025 (Fuji Film). \color{black}


\subsection{Experimental results}
\label{sec:PositronAcceleration}
\begin{figure}[h!]
\centering
\includegraphics[width=0.4\textwidth]{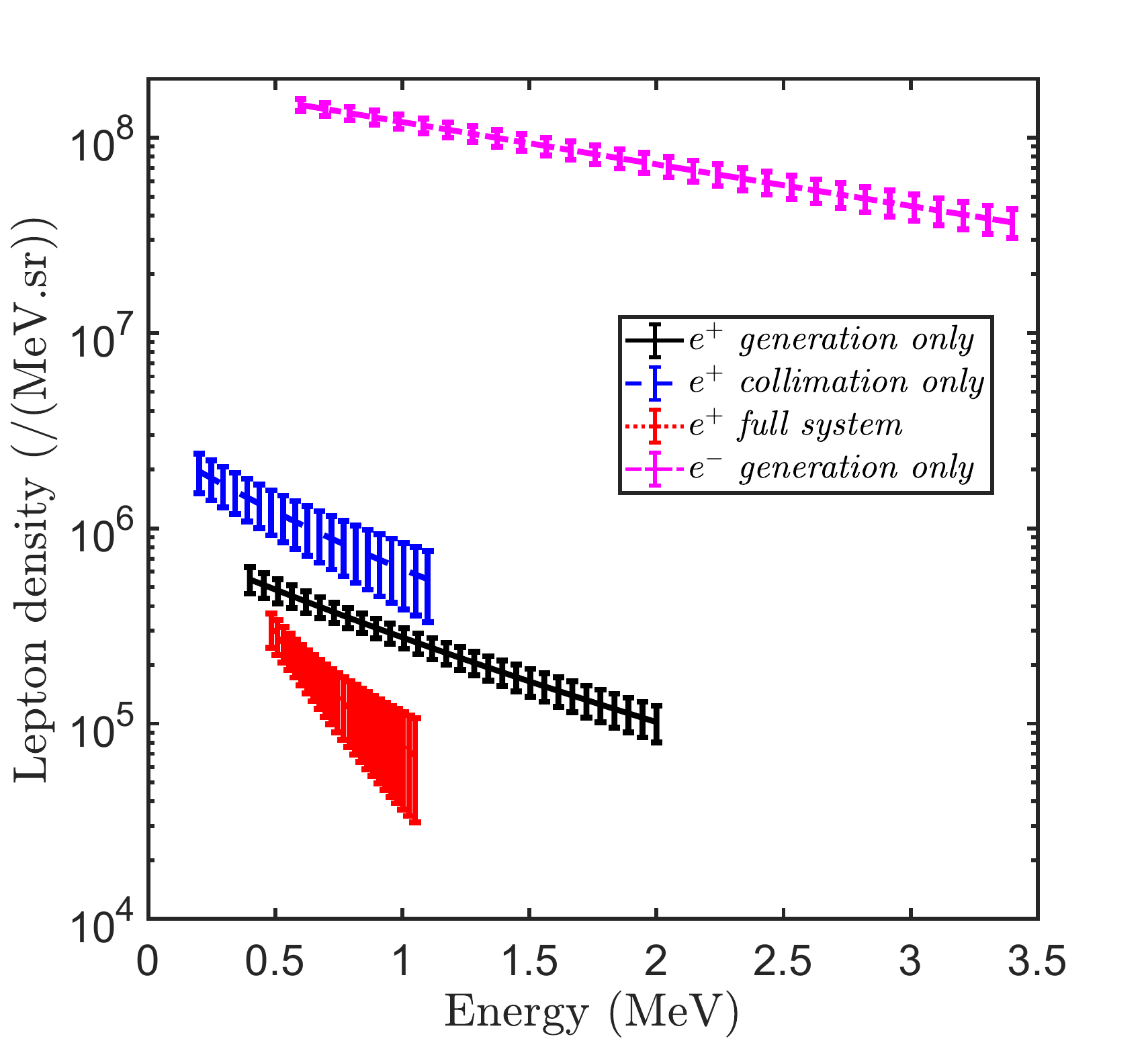}
\caption{Experimental positron spectra obtained using the full system (red dotted line), the quadrupole doublet but no second dipole (blue dashed line), and using no quadrupole and no second dipole (black solid line). The mean electron spectra escaping the solid target in that last configuration is also shown for comparison (dashed dotted magenta line). Error bars represent the standard deviation in each data set. \color{black}}
\label{fig:experimentalSpectra}
\vspace{5mm}
\end{figure}
Fig.~\ref{fig:experimentalSpectra} shows the positron spectra per steradian averaged over several shots in the three different configurations, with error bars representing the standard deviation in each data set \color{black}; as it can be seen, the positron generation was fairly stable throughout the experiment. 
The spectra all exhibit an exponentially decreasing profile and the total number of detected positrons in the different configurations were $N^{e^+}_{generation} \simeq 3.1\times10^3$, $N^{e^+}_{collimation} \simeq 7.2\times10^3$ and $N^{e^+}_{full} \simeq 4.2\times10^2$ respectively. Electrons obtained during the same shots as the positrons with no quadrupoles and without the second dipole are also shown as a dashed-dotted magenta line in Fig.~\ref{fig:experimentalSpectra}. The electron yield is approximately two orders of magnitude higher than the positron yield, as expected in this configuration \cite{Sarri:2015aa}.

\begin{figure}[b!]
\centering
\includegraphics[width=0.8\textwidth]{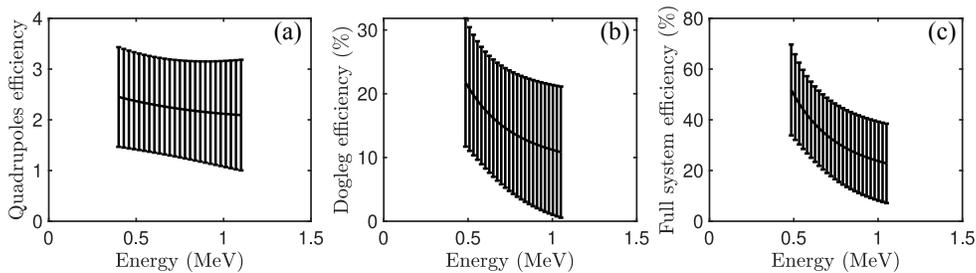}
\caption{(a) Ratio between the positron with (blue dashed line in Fig. \ref{fig:experimentalSpectra}) and without (black solid line in Fig. \ref{fig:experimentalSpectra}) the Hallbach magnets.
(b) Ratio between the positron spectrum with (red dotted line in Fig. \ref{fig:experimentalSpectra}) and without (blue dashed line in Fig. \ref{fig:experimentalSpectra}) the second dipole magnet in the dogleg. 
(c) Ratio between the  positron spectrum after the whole system (red dotted line in Fig. \ref{fig:experimentalSpectra}) and the positron spectrum recorded without second dipole or Hallbach magnets (black solid line in Fig. \ref{fig:experimentalSpectra}).
}
\label{fig:quadDoglegEfficency}
\vspace{5mm}
\end{figure}

The performance of the quadrupoles and dogleg is exemplified in Fig.~\ref{fig:quadDoglegEfficency}(a) as a function of energy, limited to the common energy range of 0.5 - 1 MeV; the addition of the quadrupole doublet leads to an increase of the detected positrons by more than a factor of 2. This, when combined with the efficiency of the dogleg,
implies an energy-dependent efficiency of the entire system of the order of 30 - 50 \% (Fig.~\ref{fig:quadDoglegEfficency}(c)). \color{black}


\subsection{Numerical Modelling}
\label{sec:NumericalModelling}

\begin{figure}[h!]
\centering
\includegraphics[width=1\textwidth]{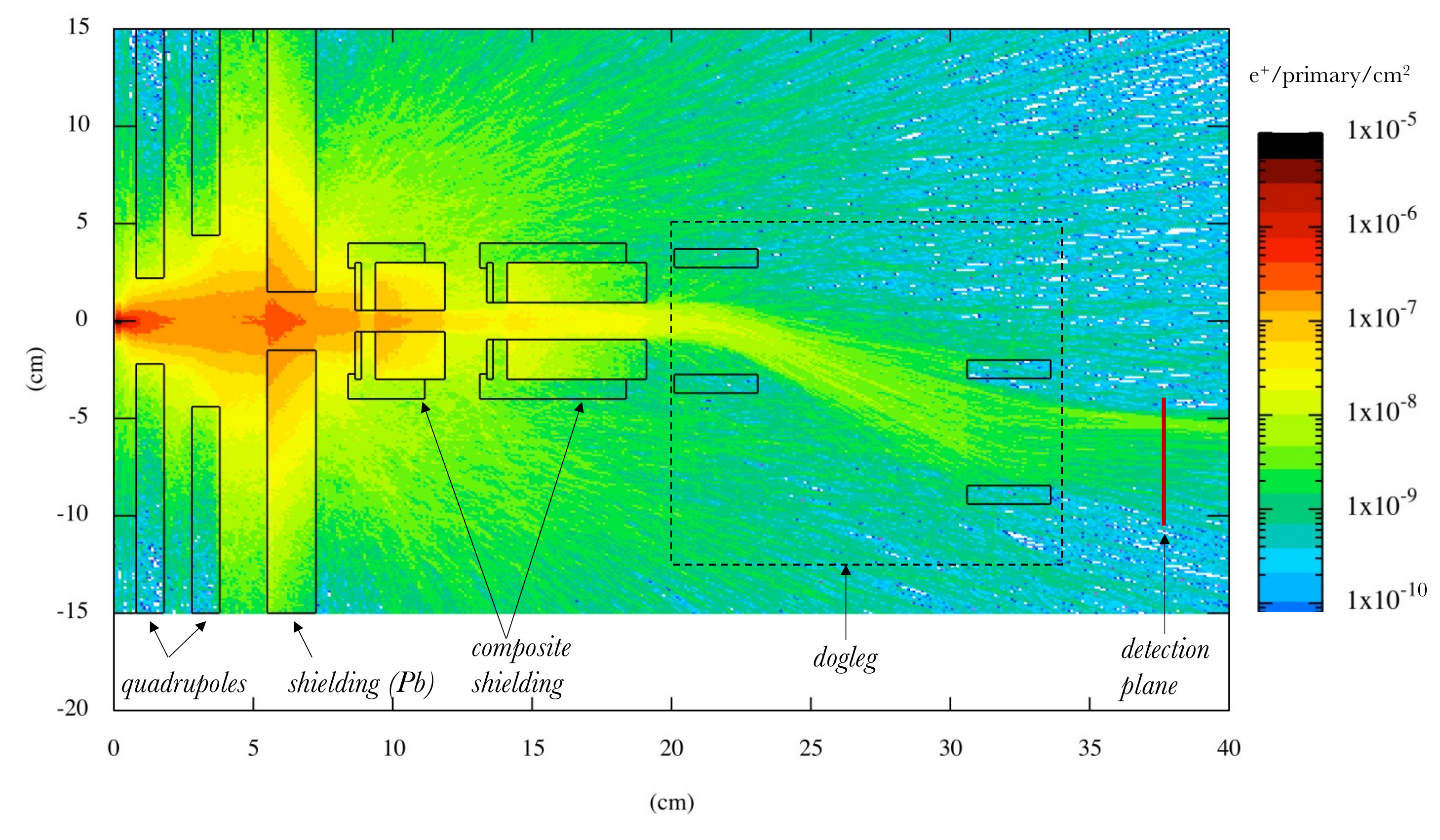}
\caption{Top-view of the spatial distribution of the positron beam through the whole system, as simulated by the Monte-Carlo scattering code FLUKA. The colorbar is in logarithmic scale and it is in units of positrons per primary electron per cm$^2$.\color{black}}
\label{usrbin}
\vspace{5mm}
\end{figure}

\begin{figure}[b!]
\centering
\includegraphics[width=1\textwidth]{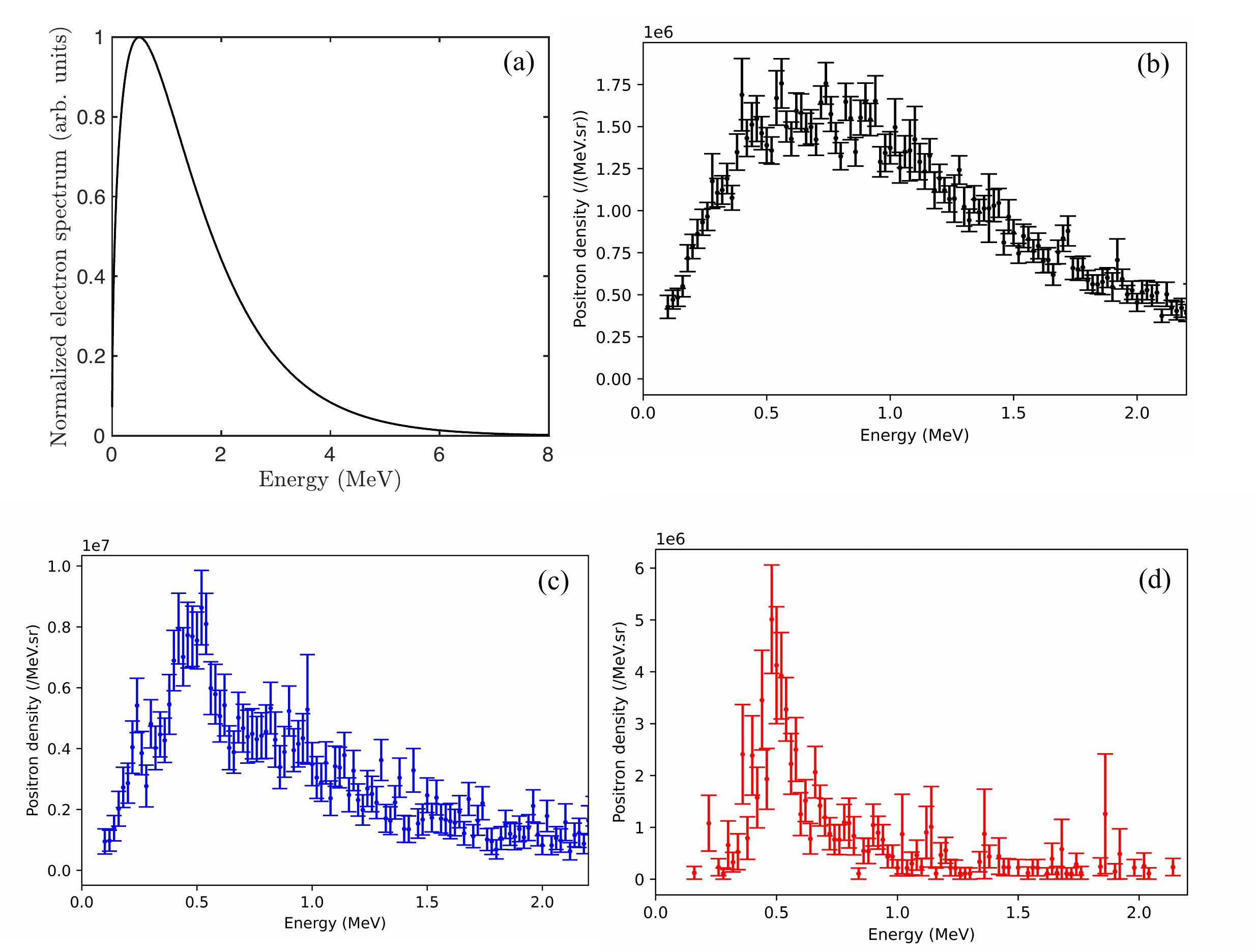}
\caption{(a) : Normalized electron energy distribution used as input for FLUKA simulations. (b) : Positron spectra obtained with FLUKA in the \textit{generation only} configuration without quadrupoles or second dipole; (c) in the \textit{collimation only} with quadrupoles but no second dipole and (d) using the full system.} 
\label{fig:flukaSpectra}
\vspace{5mm}
\end{figure}

In order to numerically confirm the experimental results presented in the previous section, modelling of the experiment was performed using the Monte-Carlo code FLUKA ~\cite{flukaa,BOHLEN2014211}.
An electron population with a Maxwellian distribution and an electron temperature of 0.9 MeV was chosen as an input for the simulation (Fig.~\ref{fig:flukaSpectra}(a)), in agreement with the experimental results using a thin gold target with a thickness of $50\mu$m (Fig.~\ref{el_spectrum}). The simulations were performed with $4\times 10^9$ primary electrons and are scaled up to $5 \times10^{10}$ primaries for a direct comparison with experimental results. Due to computational constraints, a pencil-like electron beam with zero temporal duration and a point-like source was assumed. 

As an example of the expected performance of the whole system we plot in Fig. \ref{usrbin}  a simulated top-view of the spatial distribution of the positrons after the 2mm Ta converter target. As one can see the majority of the positrons are collimated by the two quadrupoles into the shielding and dog-leg, resulting in a sizeable population of positrons at the detector plane.  \color{black}

\begin{figure}[h!]
\centering
\includegraphics[width=1\textwidth]{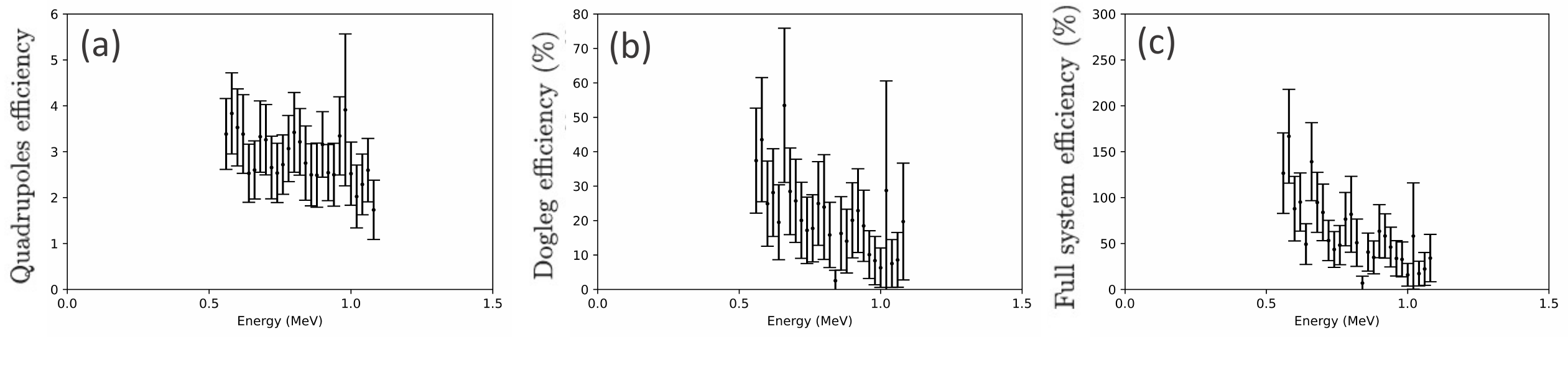}
\caption{Simulated results of the (a) the ratio between the positron with (blue dashed line in Fig. \ref{fig:experimentalSpectra}) and without (black solid line in Fig. \ref{fig:experimentalSpectra}) the Hallbach magnets,
(b) the ratio between the positron spectrum with (red dotted line in Fig. \ref{fig:experimentalSpectra}) and without (blue dashed line in Fig. \ref{fig:experimentalSpectra}) the second dipole magnet in the dogleg, and 
(c) the ratio between the  positron spectrum after the whole system (red dotted line in Fig. \ref{fig:experimentalSpectra}) and the positron spectrum recorded without second dipole or Hallbach magnets (black solid line in Fig. \ref{fig:experimentalSpectra}).
}
\label{efficiency_sim}
\vspace{5mm}
\end{figure}

All previously described configurations were simulated for a detailed comparison with the experiment. In all cases, the magnetic dipoles are modelled by assuming a magnetic field with a super-Gaussian distribution with a peak field of 53 mT and a width of 15 mm, reproducing the measured field maps. \color{black}
The quadrupoles are also simulated assuming magnetic field gradients corresponding to the experimental values.\\
Fig.~\ref{fig:flukaSpectra}(b-d) shows the positron spectra obtained with FLUKA in the three different configurations. The simulated spectra qualitatively agree with the experimental values but show a 3-5 times higher yield in all the three different cases. This is to be attributed to the assumption of a pencil-like electron beam in the simulation, which results in a higher positron population to be guided through the system. This overestimate of the positron yield, however, does not affect the accuracy of the simulated efficiency of the dogleg and of the quadrupoles (shown in Fig. \ref{efficiency_sim}), which are in good agreement with the experimental values. \color{black}

\begin{figure}[h]
\centering
\includegraphics[width=0.6\textwidth]{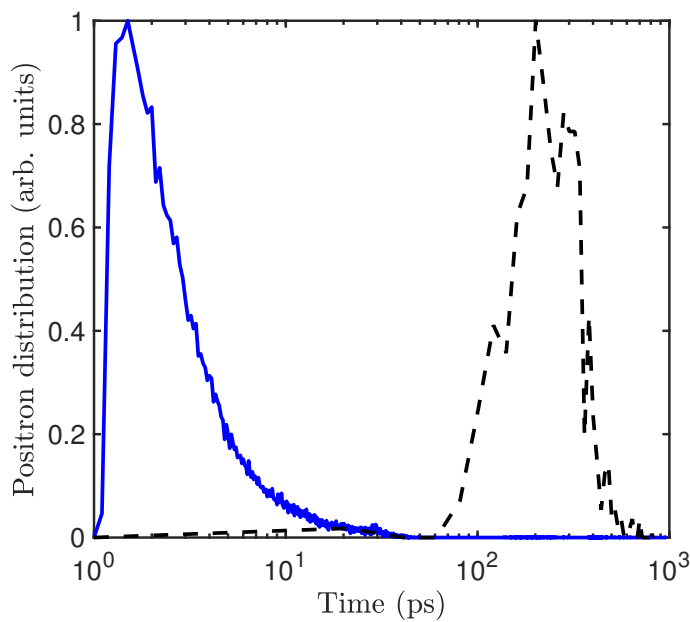}
\caption{Positron temporal distributions obtained with FLUKA at the target back surface (blue solid line) and after the dogleg (dashed black line).}
\label{fig:positronsTime1}
\end{figure}

The FLUKA simulations, in good agreement with the experimental values, indicate an energy-dependent dogleg efficiency ranging from $~30 \%$ at $0.5$ MeV to $~10\%$ at $1$ MeV and an efficiency of the Hallbach magnets between 2 and 3.\\ 
Using a custom fortran routine, the particles time of arrival was also scored during these simulations and is displayed in Fig.~\ref{fig:positronsTime1}. The blue solid line shows the temporal distribution of the positrons as they exit the target whereas the black dashed line shows their distribution after the dogleg. The positron beam exhibit a $\nicefrac{1}{e^2}$ time duration of $\tau_{1/e^2}^{target} \simeq 5$ ps after the target and $\tau_{1/e^2}^{dogleg} \simeq 340$ ps after the dogleg. It must be noted that these results do not take into account the duration of the primary electron beam at source, which can be estimated as $\tau_e\simeq 1.2 \tau_l \simeq 1$ ps \cite{Fuchs:2006aa} and should be added to the results reported here.

The longer duration of the positron beam at the detector plane is due to their broad spectrum. This is because positrons of different energy will have a different time of flight between the target and the detection plane after the dogleg, as well as different trajectories through the dogleg: these are responsible for the temporal lengthening of the positron beam. 
The temporal distributions shown in Fig.~\ref{fig:positronsTime1} correspond to the entire positron beam; any energy selection within the system will result in a smaller energy spread and, therefore, a shorter positron duration at the detector plane. This is because a narrower energy spread would restrict the different paths through the system as well as the temporal spreading due to the different time of flight of positrons with different energies. As we will discuss in more detail in the following section, selecting an energy slice of 50 keV leads to a positron beam FWHM duration of 50 - 60 ps, whilst maintaining a sizeable number of positrons per second. \color{black}

\section{Extension to different laser system : laser wakefield electrons conversion}
\label{sec:Extension}

Even though the results in Sec.~\ref{sec:PositronAcceleration} \& \ref{sec:NumericalModelling} already demonstrate 
interesting positron properties, their use for practical applications would be severely affected by the low number of positrons per second that a typically low repetition rate Nd:glass high-energy laser systems would provide. As an example, TARANIS can only operate at a maximum of a shot every 15 minutes. 
In the following, these results will then be applied to a different approach. This is motivated by the recent availability of TW-scale laser systems with ultra-short (of the order of few fs) pulse duration, and kHz repetition rate. The ultra-short pulse duration of these systems allows them to drive a laser-wakefield accelerator (LWFA) able to generate fs-scale electron bunches \cite{RevModPhys.81.1229,Tajima:1979aa}. 
In the following, we show that positron beams with a high flux per second and a short duration per energy slice can be generated using this class of laser systems. This is thanks to the increase in repetition rate to the kHz level and the higher electron energy accessible ($\gtrsim1$ MeV, see for example Ref. \cite{Guenot:2017aa}), amply compensating for the lower charge of LWFA electron bunches (typically of the order of tens to few hundreds of pC). \color{black}

We will use, as an example, the SYLOS2 laser system which is operational at ELI-ALPS Research Institute in Hungary to study the characteristics of the positron beam such a system could achieve. The SYLOS2 is a 1 kHz-repetition-rate, 4.8 TW optical parametric chirped pulse amplification (OPCPA) laser system that has demonstrated stable long-term operation at 32 mJ output energy and 6.6 fs laser pulse duration at 900 nm central wavelength \cite{Toth:2020aa}.

\subsection{Simulations of LWFA electron beam}
\begin{figure}[h]
\centering
\includegraphics[width=8cm]{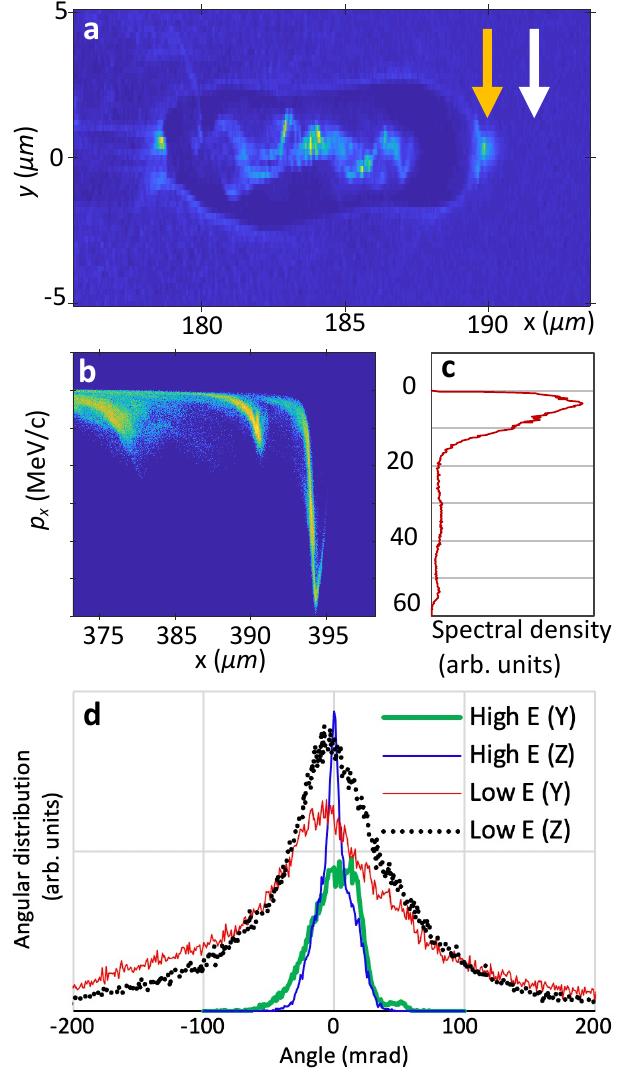}
\caption{(a) Plasma density profile snapshot after transition to a beam-driven wake, with the orange arrow showing the position of the driving electron bunch, and the white one showing the position of the laser pulse. (b) longitudinal phase-space plot of the electrons and (c) corresponding spectrum. (d) Angular distribution of the high- and low-energy electron population in both transverse directions.}
\label{fig:SYLOS_ebeam}
\vspace{5mm}
\end{figure}
Numerical simulation of the acceleration was carried out using the EPOCH3D Particle-in-Cell code \cite{Arber:2015aa}. The simulation domain was a 24 $\mu $m $ \times $ 30 $\mu $m $ \times $ 30 $\mu $m moving window with free boundaries, and a mesh resolution on 50 nm $\times$ 200 nm $\times$ 200 nm, with 2 particles per cell. The laser pulse parameters - as expected on-target from the SYLOS2 laser system - were 28 mJ, 7 fs, at 900 nm wavelength, focused to a $2.2 ~\mu m$ FWHM focal spot size for the maximum vacuum intensity of $3 \times 10^{19}$ $Wcm^{-2}$ and $a_0 = 4.3$, propagating in the $x$ direction.
The simulated target was pure $N_2$ gas with a supergaussian profile of order $2.8$, with a density "plateau" of 100 $\mu m$ between $90\%$ density values and 100 $\mu$m ramps (between $10\%-90\%$ density values). The laser focused at the start of the plateau (at the first $90\%$ density value). The background electrons from the Nitrogen L-shell were assumed to be pre-ionized, while the two K-shell electrons were non-ionized. Ionization injection was modelled using the EPOCH built-in routines for field, barrier suppression and multiphoton ionization processes. The background electron density of the target was $6\times 10^{19}$ cm$^{-3}$ corresponding to a plasma wavelength of 4.3 $\mu$m.

Fig.~\ref{fig:SYLOS_ebeam}(a) shows the plasma bubble at the end of the density plateau. At this stage the laser beam is depleted, with $a_0 \sim 0.2$ for the driving pulse (position shown by white arrow), and the wake is now driven by the first injected electron bunch (orange arrow) due to the very high beam loading. The plasma bubble is significantly elongated, resulting in a $\sim 10 \mu$m long electron bunch length from the continuous injection. Fig.~\ref{fig:SYLOS_ebeam}(b) shows the longitudinal phase-space of the accelerated electron beam after exiting the plasma (180 $\mu$m after the plateau end). The high-energy (above 18 MeV) tail of the spectrum is entirely from the leading edge of the electron beam. It extends to 60 MeV and has an average energy of 37 MeV and 27 pC of total charge ($1.7\times 10^8$ electrons). The low-energy spectrum, with a peak at 3.5 MeV, is from the continuous injection following the leading bunch, with a total charge of 137 pC ($8.6\times10^8$ electrons).
The angular divergence of the electron beam is shown in Fig.~\ref{fig:SYLOS_ebeam}(d), the low-energy fraction has a FWHM divergence of 115 mrad in the $y$ (laser polarization) direction and 78 mrad in the $z$ direction. The high-energy fraction has a smaller divergence, 41 mrad (14 mrad) FWHM in the $y$ ($z$) direction.

\subsection{Conversion of LWFA electron beam to positrons}

\begin{figure}[ht]
\centering
\includegraphics[width=1\textwidth]{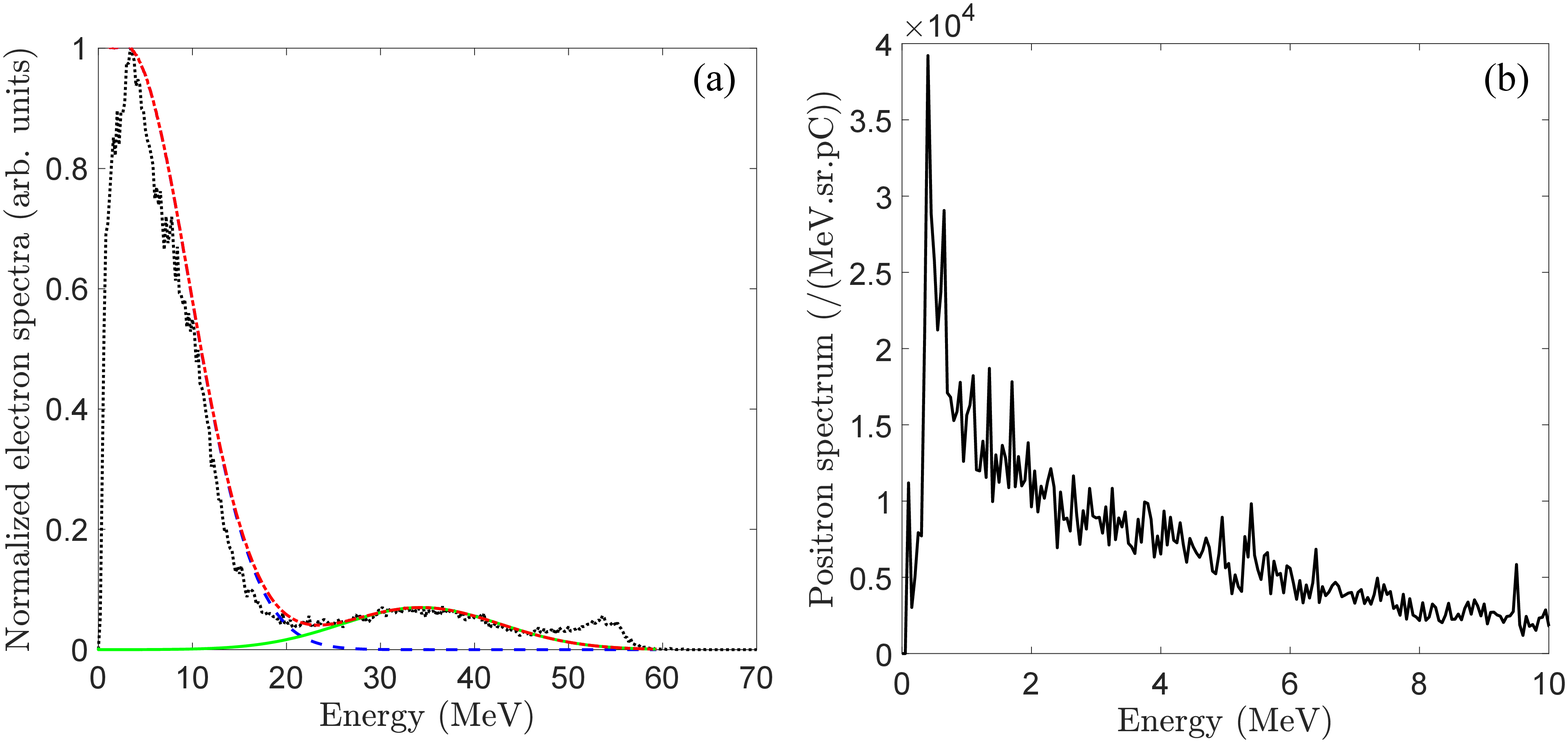}
\caption{(a) Normalized electron spectra : electron energy distribution predicted by the PIC simulation (black dotted line), low-energy (blue dashed line) and high-energy (green solid line) fitting of the electron distribution, and their sum of (red dotted dashed line). (b) Resulting simulated positron spectrum at the detector plane, per steradian and per pC of initial electron charge.}
\label{fig:sylos2FlukaSpectra}
\vspace{5mm}
\end{figure}

\begin{figure}[ht]
\centering
\includegraphics[width=1\textwidth]{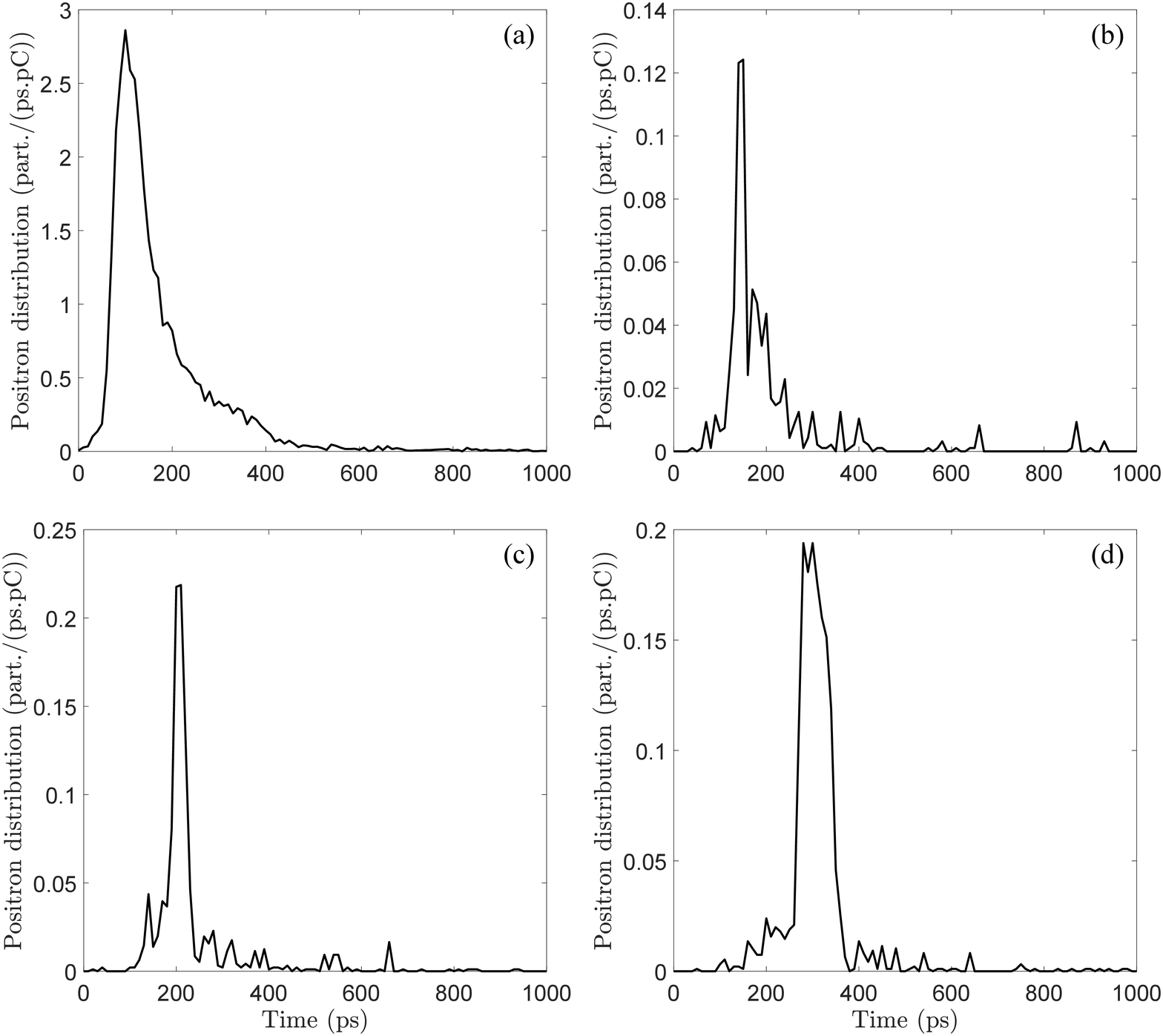}
\caption{Temporal distribution of the positrons after the dogleg: for the whole spectrum (a), for positrons with an energy of $1\pm0.05$ MeV (b), for positrons with an energy of $700\pm50$ keV (c), for positrons with an energy of $500\pm50$ keV (d).}
\label{fig:sylos2FlukaTime}
\vspace{5mm}
\end{figure}

A suitable fitting of the electron spectrum predicted by PIC simulations (see Fig.~\ref{fig:sylos2FlukaSpectra}(a)) was used as an input for a FLUKA simulation of the \textit{full system}, i.e., including the 2mm tantalum converter target, the quadrupole magnets, the collimators, and the dogleg, as shown in Sections II and III.

In order to account for the different divergence of the low-energy (below $\sim18$ MeV) and high-energy (above $\sim 18$ MeV) components of the electron spectrum, two different simulations were performed and combined together. The input of the first simulation was an electron beam with a low energy distribution exhibiting a gaussian angular distribution with a 100 mrad FWHM width, displayed as a blue dashed line on Fig.~\ref{fig:sylos2FlukaSpectra}(a). In a second simulation, the electron beam input, shown on Fig.~\ref{fig:sylos2FlukaSpectra}(a) as a solid green line, consisted in a Gaussian energy distribution centered on 35 MeV with a 20 MeV FWHM energy spread and a gaussian divergence of 25 mrad FWHM. Combining and scaling these two simulations resulted in the initial electron distribution shown in Fig.~\ref{fig:sylos2FlukaSpectra}(a) as a red dotted dashed line. 

The interaction of such an electron beam with a 2mm thick Ta target generated a positron beam which, after collimation and propagation through the magnetic system, resulted in the distribution shown in Fig.~\ref{fig:sylos2FlukaSpectra}(b) recorded at the exit of the dogleg. Similarly to what was seen in Sec.~\ref{sec:NumericalModelling}, the positron distribution after the dogleg is peaked around 500 keV as a consequence of the higher efficiency of the system for this energy. 

Fig.~\ref{fig:sylos2FlukaTime} shows the temporal distribution of the positron beam after the dogleg for different energy slices\color{black}. The total positron population has a temporal distribution with a FWHM of 135 ps. As mentioned earlier, restricting the allowed energy slice  (for instance, with a slit within the dogleg) \color{black} results in a shorter positron beam duration. As an example, allowing only an energy slice \color{black} of $\pm$ 50 keV reduces the temporal duration down to a FWHM of 50 - 60 ps. A detailed summary of the positron bunch duration obtained in different energy slices is given in Fig. ~\ref{fig:sylos2FlukaTime} and Table I. Even in a 50 keV energy slice\color{black}, a realistic primary electron beam conservatively containing 10 pC of charge will approximately produce 100 positrons per shot. Operating at a 1 kHz repetition rate, this would then translate into more than $10^5$ positrons per second in a 50 keV energy slice \color{black}, well within the requirements for PALS. 

%
%

\begin{table}
\begin{center}
\begin{tabular}{|c|c|c|c|}
\hline 
Energy slice & FWHM (ps) & $N_{e^{+}}$ per pC of $e^-$ &  Positron flux for 10 pC $e^-$ at 1 kHz ($e^+$/s) \\
\hline 
\hline
Whole distribution & 135 & $\sim 338.4$ & $\sim 3.4\times10^{6}$ \\
\hline 
$1\pm0.05$ MeV & 50 & $\sim 7.4$ & $\sim 7.4\times10^{4}$ \\
\hline 
$700\pm50$ keV & 60 & $\sim 11.0$ & $\sim 1.1\times10^{5}$ \\
\hline 
$500\pm50$ keV & 50 & $\sim 16.5$ & $\sim 1.6\times10^{5}$ \\
\hline 
\end{tabular}
\caption{Temporal duration and number of positrons after the dogleg for different energy slices. The positron flux is given assuming a 10 pC primary electron bunch at a 1 kHz repetition rate.}
\label{tab:flukaTimeDurationsNumbers}
\end{center}
\end{table}
\color{black}

\begin{figure}[t!]
\centering
\includegraphics[width=0.5\textwidth]{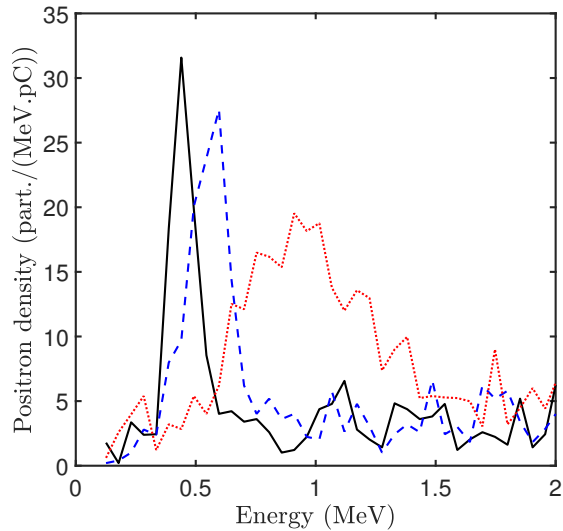}
\caption{Example of energy selection after the dogleg. All distributions correspond to a 4.9 mm wide position selection centered on $\sim 72.0$ mm (black solid line), $\sim 55.9$ mm (blue dashed line), $\sim 34.6$ mm (red dotted line) from the main axis.}
\label{fig:sylos2FlukaEnergySelection}
\vspace{5mm}
\end{figure}

Due to the transverse spatial chirp induced by the dogleg on the positron beam, on-shot energy selection can be easily achieved by introducing a moveable slit after the dogleg. This is exemplified in Fig.~\ref{fig:sylos2FlukaEnergySelection}, which shows the positron energy distribution at different points along the transverse axis after the dogleg. The position-energy correlation introduced by the dogleg thus allows for energy selection by selecting the position of the slit. In practice, a slit would be placed just after the dogleg to select the required part of the positron spectrum and shield the rest of the beam. In the example shown in Fig.~\ref{fig:sylos2FlukaEnergySelection}, the virtual slit is 4.9 mm wide and select positrons with energies (peak $\pm$ FWHM) of $\sim 440 \pm 130$ keV (black solid line), $\sim 600 \pm 200$ keV (blue dashed line) and $\sim 910 \pm 630$ keV (red dotted line). \\
With this setup, the energy selection can be adjusted with the position of the slit and the energy spread can be adjusted with the width of the slit at the expense of the number of positrons reaching the sample. Furthermore, the quadrupoles and dipoles fields could be adjusted to allow a different energy band to go through the dogleg and be selected in the same fashion.

\section{Discussion and conclusions}
\label{sec:Conclusion}
In summary, we report on experimental and numerical studies demonstrating the suitability of laser-driven positron beams as probes for high-resolution and volumetric scanning of materials. Preliminary experiments using the TARANIS laser and a compact beam-line already indicate good efficiency in collection and energy-selection of positrons generated during the interaction of a laser-driven electron beam with a thick tantalum target. Numerically extending these results to the next generation of high repetition rate low-energy laser systems indicates that more than $10^5$ positrons in a 50 keV energy slice per second can reach the sample to be probed, with an energy tuneable virtually from a fraction up to a few MeV and a FWHM duration of the order of 50 ps. A comparison between the results reported in this work and the performance of existing machines devoted to PALS is reported in table \ref{comparison}. As one can see, a positron flux of approximately $10^5$ $e^+/s$ is in line with other systems, with the extra advantages of a significantly shorter positron bunch duration and a tuneability over a larger energy range, extending up to the MeV scale.\color{black}


\begin{table}[h!]
\begin{center}
\begin{tabular}{|l|c|c|c|}
\hline 

Positron source & Positron flux ($e^+$/s) & Beam duration FWHM (ps) & Positron energy (keV) \\
\hline 
\hline
Our work & $10^5$ - $10^6$ & 50 - 60 & 0 - $10^3$ \\
\hline 
ELBE \cite{sponsor} & $10^6$ & 250 & 0.5 - 15 \\
\hline 
NEPOMUC \cite{nepomuc} & $10^9$ & / & 1 \\
\hline 
PLEPS  \cite{pleps} & 0.5 - 1$\times10^4$ & 260-280 & 0.5 - 20 \\
\hline 
PALS-200A (Fuji) \cite{Fuji} & 5$\times10^2$ & 300 & 0.5 - 15 \\
\hline 
NANOPOS \cite{NANOPOS} & $10^5$ & / & 0.25 - 25 \\
\hline 
PULSTAR \cite{PULSTAR} & $10^6$ - $10^9$ & 300 & 0.5 - 10 \\
\hline 
\end{tabular}
\caption{Comparison of positron parameters obtainable at SYLOS2 with other positron facilities dedicated to PALS.}
\label{comparison}
\end{center}
\end{table}


\section{Acknowledgements}
We acknowledge support from the Engineering and Physical Sciences Research Council (grant numbers: EP/V044397/1, EP/N027175/1, EP/P010059/1, and EP/T021659/1).

Four of the authors (D.P., S.L., C.K., N.A.M.H.) are supported by the European Union through the ELI-ALPS Project under Grant GINOP-2.3.6-15-2015-00001 and in part by Horizon 2020, the EU Framework Programme for Research and Innovation under Grant Agreement No. 654148 and No. 871124 Laserlab-Europe. 
N. A. M. H. acknowledges the President International Fellowship Initiative (PIFI) of the Chinese Academy of Sciences; the International Partnership Program (181231KYSB20170022) of CAS; the Inter-Governmental Science and Technology Cooperation of MOST.

\end{document}